\documentclass[aps,twocolumn,amsmath,amssymb,nofootinbib,preprintnumbers,showpacs]{revtex4}
\usepackage{epsfig} \usepackage[x11names]{xcolor}
\usepackage{mathrsfs,graphicx,bm,amsmath}

\definecolor{jan}{rgb}{0.0,0.7,0.0}
\definecolor{holger}{rgb}{0.5,0,0.5}
\definecolor{christof}{rgb}{0.8,0,0}
\newcommand{\STr}{\mathrm{STr}}

\newcommand{\kF}{k_{\text{F}}}

\newcommand{\sigex}{\tilde{\sigma}}

\newcommand{\Tn}{T}

\def\di{\displaystyle}

\def\bg{\begin{eqnarray}\begin{array}{rcl}\displaystyle}
\def\eg{\end{array} &\di    &\di   \end{eqnarray}}
\def\bm#1{\begin{eqnarray}\begin{array}{#1}\di}
\def\bmo#1{\begin{eqnarray*}\begin{array}{#1}\di}
\def\bml#1#2{\begin{eqnarray}\begin{array}{#1}\label{#2}\di}
\def\bgo{\begin{eqnarray*}\begin{array}{rcl}\displaystyle}
\def\ego{\end{array} &\di    &\di \nonumber  \end{eqnarray*}}

\def\btensor#1#2{\renew\left#1\begin{array}{#2}\di}
\def\brtensor#1#2#3{\ren#3\left#1\begin{array}{#2}}
\def\botensor#1#2{\renew\left#1\begin{array}{#2}}
\def\etensor#1{\end{array}\right#1}

\def\Eq#1{Eq.~(\ref{#1})}

\def\STr{{\rm STr}}

\def\s0#1#2{\mbox{\small{$ \frac{#1}{#2} $}}}
\def\0#1#2{\frac{#1}{#2}}

\begin{document}

\preprint{HD-THEP-07-02}

\title{Flow Equations for the BCS-BEC Crossover}

\author{S. Diehl${}^{a,b}$}
\author{H.Gies${}^{a}$}
\author{J. M. Pawlowski${}^{a}$}
\author{C. Wetterich${}^{a}$}

\affiliation{\mbox{\it ${}^a$Institut f{\"u}r Theoretische Physik,
Philosophenweg 16, D-69120 Heidelberg, Germany}\\
\mbox{\it ${}^b$Institute for Quantum Optics and Quantum
Information of the Austrian Academy of Sciences,}\\
\mbox{\it A-6020 Innsbruck, Austria}}



\begin{abstract}
  The functional renormalisation group is used for the BCS-BEC
  crossover in gases of ultracold fermionic atoms. In a simple
  truncation, we see how universality and an effective theory with
  composite bosonic di-atom states emerge. We obtain a unified picture
  of the whole phase diagram. The flow reflects different effective
  physics at different scales. In the BEC limit as well as near the
  critical temperature, it describes an interacting bosonic theory.
\end{abstract}


\maketitle


Ultracold gases of fermionic atoms near a Feshbach resonance show a
crossover \cite{p5ALeggett80} between Bose-Einstein condensation (BEC)
of molecules and BCS superfluidity. The controlled microphysics, which
can be measured by two-body scattering and the molecular binding
energy, and recent experimental breakthroughs \cite{Exp} can open a
new field of quantitatively precise understanding of complex many body
physics. On the theory side, this calls for a quantitative and reliable
approach to strongly interacting systems. In turn, a precise
experimental control of the relevant parameters, namely the scattering
length $a(B)$ depending on the magnetic field $B$, the density $n$ and
the temperature $T$, can test the viability of non-perturbative
methods.

The functional renormalisation group (FRG) directly connects the
'microphysics' to observable 'macrophysics' by a non-perturbative flow
equation \cite{Wetterich:1992yh}. It has been used successfully for
precision estimates in simple non-perturbative systems and has already
been applied to coupled systems of fermions and collective bosonic
degrees of freedom in relativistic \cite{Berges:1997eu,Gies:2001nw}
and non-relativistic theories \cite{Birse:2004ha,Baier:2003fw}. In
this approach, the results of perturbative renormalisation near the
critical dimension \cite{Nishida:2006br} or for a large number of
components $N$ \cite{NS} can be recovered by an appropriate level of
truncation of an exact functional differential equation. In a certain
sense, the FRG can be regarded as a differential form of
Schwinger-Dyson or gap equations in a 1PI \cite{Diehl:2005ae} or 2PI
\cite{Rantner} setting, see \cite{Pawlowski:2005xe}.

\emph{Method and approximation scheme} -- We
study the scale dependence of the average action $\Gamma_k$
\cite{Berges:2000ew}. It includes all quantum and
thermal fluctuations with momenta $q^2\gtrsim k^2$, or in the presence
of a Fermi surface with effective chemical potential $\sigma>0$, all
$|q^2-\sigma|\gtrsim k^2$. For $k\to 0$, all fluctuations are included
and $\Gamma_{k\to 0}$ generates the 1PI correlation functions. In
practice, this is realised by introducing suitable cutoff functions
$R_k(q)$ in the inverse propagators. The dependence of $\Gamma_k$ on
$k$ obeys an exact flow equation \cite{Wetterich:1992yh},
\begin{eqnarray}\label{eq:FRG}
  \partial_k \Gamma_k &=& \frac{1}{2} \STr \,
  (\Gamma^{(2)}_k + R_k)^{-1}\,\partial_k R_k .
\end{eqnarray}
Here, $\STr$ sums over spatial momenta $\vec q$ and Matsubara
frequencies $\omega_{\text{M}}$ as well as over internal indices and
species of fields, with a minus sign for fermions.  The second
functional derivative $\Gamma_k^{(2)}$ represents the full inverse
propagator in the presence of the scale $k$. Both $\Gamma_k$ and
$\Gamma_k^{(2)}$ are functionals of the fields.

In the present Letter, we demonstrate that already a very simple
truncation of $\Gamma_k$ is sufficient to account for all qualitative
features and limits of the crossover problem. We approximately solve
Eq.~\eqref{eq:FRG} with the ansatz
\begin{eqnarray}\label{eq:Trunc}
  \Gamma_k \hspace{-0.15cm}&=&
  \int_T  d^4 x \Big[\psi^\dagger\big(\partial_{\tau}
  -\triangle -\sigma\big)\psi
+ \varphi^*\big(\partial_\tau - A_\varphi\triangle\big)
  \varphi
   \nonumber\\
  &&\qquad\quad + u(\varphi) - h_\varphi\Big({\varphi}^*\psi_1\psi_2
  - {\varphi}\psi^*_1\psi^*_2\Big)\Big].
\end{eqnarray}
In addition to the fermionic fields $\psi$ for the open-channel atoms,
we use a collective bosonic di-atom field $\varphi$. Depending on the
region of the phase diagram and the scale $k$, it can be associated
with microscopic molecules, Cooper pairs, effective macroscopic bound
states or simply represents an auxiliary field. The bosonic field is
renormalised by a wave function renormalisation,
$\varphi=Z_\varphi^{1/2} \hat{\varphi}$, such that at every scale $k$
the term linear in the Euclidean time derivative $\partial_\tau$ has a
standard normalisation. (For the fermions, this
renormalisation is omitted.) Eq. \eqref{eq:FRG} holds for fixed
unrenormalised fields $\hat\varphi$, i.e.,
$(\Gamma_{k,\varphi}^{(2)})_{\alpha\beta}=\partial^2
\Gamma_k/\partial\hat{\varphi}_\alpha \partial\hat{\varphi}_\beta$. We
define \cite{Diehl:2005ae}
\begin{equation}
  Z_\varphi=-\frac{\partial
    \Gamma_{k,\varphi}^{(2)}(\omega,\vec{q}=0)}{\partial\omega}
  \Big|_{\omega=0}, \label{eqE1A}
\end{equation}
where $\Gamma^{(2)}_k$ is evaluated for an analytically continued
Matsubara frequency $\omega_{\text{M}}\to i \omega$. The fields
and couplings in Eq.~\eqref{eq:Trunc} are scaled with powers of an
appropriate momentum scale $\hat{k}$ or energy scale $\hat{k}^2/(2M)$
\cite{Diehl:2005ae}. For nonzero density $n$, we choose the Fermi
momentum $\hat{k}=\kF=(3\pi^2 n)^{1/3}$. Our units are $\hbar = c =
k_B =1$.

We consider a polynomial effective potential $u(\varphi)$ written in
terms of $\rho = \varphi^*\varphi$,
\begin{eqnarray}\label{eq:u}
  u &=& \left\{
  \begin{array}{ll}
    m_\varphi^2 \rho +   \frac{1}{2} \lambda_\varphi \rho^2\quad &
    \text{SYM} \\
     \frac{1}{2} \lambda_\varphi (\rho-\rho_0)^2 \quad & \text{SSB}
  \end{array}.
  \right.
\end{eqnarray}
Here, we distinguish the symmetric regime (SYM), where the minimum of
$u$ is at $\rho=0$ and $m_\varphi^2\geq0$, from the regime with
spontaneous breaking of the U(1) symmetry (SSB), where the potential
minimum occurs at $\rho_0(k)$. Superfluidity is signalled by
$\rho_0(k\to 0)>0$, with a gap for single fermionic atoms
$\Delta=h_\varphi \sqrt{\rho_0}$.

The flow starts at some microscopic scale $k_{\text{in}}$ with
$\lambda_\varphi=0$, $m_{\varphi,\text{in}}^2>0$ and $A_\varphi=1/2$.
Here $m_{\varphi,\text{in}}^2$ is related to the magnetic field $B$
and relative magnetic moment $\mu$ by $\partial
m_{\varphi,\text{in}}^2/
\partial B =2M\mu/\hat{k}^2,$ and reflects the detuning. We will
concentrate on the limit of a broad Feshbach resonance, where
$h_{\varphi,\text{in}}^2 \to \infty$, $m_{\varphi,\text{in}}^2\to
\infty$. In this limit, the microscopic action is strictly equivalent
to a model containing only fermionic atoms with a point-like
interaction and scattering length $a$ \cite{Diehl:2005ae}. Then, the
only relevant parameter is the concentration, $c=a\kF,$ (or $a\hat{k}$
for zero density), and the Feshbach resonance is located at $a(B\to
B_0)\to \infty$. For broad resonances, the precise initial value of
$A_\varphi$ is unimportant.

Finally, we specify the regulator functions $R_k$ for fermions and
bosons. We work with optimised cutoffs
\cite{Litim:2000ci,Pawlowski:2005xe} for space-like momenta ($\xi =
q^2 - \theta (\sigma) \sigma$),
\begin{eqnarray}\label{eq:Cutoff}
  R_k^\varphi &=& Z_\varphi A_\varphi (2 k^2 - q^2) \theta (2 k^2 -
  q^2),  \\\nonumber
  R_k^\psi &=& (k^2 {\rm sgn}\, \xi- \xi) \theta (k^2 - |\xi|).
\end{eqnarray}

A central object is the flow of the effective potential $u$ with
$t=\ln k/k_{\text{in}}$, displayed here for $\sigma\leq 0$,
\begin{eqnarray}\label{PotFlow}
  \partial_t u \!&=&\! \eta_\varphi \rho u' -  \frac{k^5}{3\pi^2}
  \left(\frac{\gamma}{\gamma_\varphi}
    \tanh\gamma_\varphi -1 \right)\\\nonumber
  &&\!\!\! +\frac{2\sqrt 2 k^5}{3\pi^2}A_\varphi
  \left(\!1 - \frac{\eta_{A_\varphi}\! +\!
      \eta_\varphi}{5}\!\right)
  \left(\!\frac{\alpha\! +\!
      \chi}{\alpha_\varphi}
    \coth\alpha_\varphi -1 \!\right).
\end{eqnarray}
The functions $\gamma, \gamma_\varphi,\beta,\alpha,\alpha_\varphi,\chi$
read (for $\sigma \leq0$),
\begin{eqnarray}
  \gamma &=& \frac{ k^2 - \sigma }{2\Tn},\quad \beta =
  \frac{ h_\varphi
    \rho^{1/2}}{2 \Tn}, \quad
  \gamma_\varphi = \sqrt{\gamma^2 + \beta^2},\\\nonumber
  \alpha &=& \frac{ 2 A_\varphi k^2 +u' }{2\Tn}, \quad\chi =
  \frac{ \rho u''}{2\Tn}, \quad
  \alpha_\varphi = \sqrt{\alpha^2 + 2 \chi\alpha}.
\end{eqnarray}
Primes denote derivatives with respect to $\rho$ and the anomalous
dimensions are $\eta_\varphi = -\partial_t \ln Z_\varphi, \quad
\eta_{A_\varphi} = - \partial_t \ln A_\varphi$.  In our truncation,
the Feshbach coupling $\hat{h}_\varphi^2=Z_\varphi h_\varphi^2$ is
independent of $k$. The flow equation \eqref{PotFlow} is the analogue
of similar equations in \cite{Birse:2004ha}.

\emph{Vacuum limit} -- In order to make contact with experiment, we
have to relate the microscopic parameters to the scattering length $a$
for the two-atom scattering in vacuum. In our formalism, the vacuum
correlation functions, that directly yield the cross section
\cite{Diehl:2005ae}, are obtained from $\Gamma_{k\to0}$ in the limit
$n\to0$, $T\to0$. For fixed $\hat{k}$ the flow equations then simplify
considerably. We find that for $n=T=0$ the crossover at
finite density turns into a second-order phase transition
\cite{Diehl:2005ae,NS} as a function of $m_{\varphi,\text{in} }^2$ or
$B$, with
\begin{eqnarray}\label{VacCond}
  \begin{array}{l l l}
    { m_\varphi^2 >0, \quad \sigma_{\text{A}} = 0  }& \text{atom phase}
      & (a^{-1} < 0)  \\
    { m_\varphi^2 = 0,\quad \sigma_{\text{A}} < 0   }& \text{molecule phase}
      & (a^{-1} > 0)   \\
    { m_\varphi^2 = 0,\quad \sigma_{\text{A}} = 0  }& \text{resonance}
      & (a^{-1} = 0)
\end{array}.
\end{eqnarray}
The dimensionless ``vacuum chemical potential'' $\sigma_{\text{A}} =
\epsilon_{\text{M}}M/\hat k^2$ is related to the binding energy
$\epsilon_{\text{M}}$ of a molecule, see below.  On the BCS side, the
bosons experience a gap $m_\varphi^2>0$ and the low-density limit
describes only fermionic atoms. On the BEC side, the situation is
reversed: fermion propagation is suppressed by a gap
$-\sigma_{\text{A}}$, and the low-density limit describes bound
molecules.

In the vacuum limit, we first solve the flow equation for the mass
term $\hat m_\varphi^2 = Z_\varphi m_\varphi^2$ (we choose
$Z_{\varphi,\text{in}} =1$),
\begin{eqnarray}\label{VacFlow}
  \partial_t \hat m_\varphi^2 &=& \frac{\hat h_\varphi^2}{6\pi^2}
  \frac{k^5}{(k^2 - \sigma)^2}.
\end{eqnarray}
The condition that $\hat{m}_\varphi^2$ vanishes for $B=B_0$,
${\sigma}=0$, $k=0$ leads to
\begin{equation}
  m_{\varphi,\text{in}}^2=\hat{m}_{\varphi,\text{in}}^2=
  \frac{\hat h_\varphi^2}{6\pi^2}\,
  k_{\text{in}} + \frac{2M \mu}{\hat{k}^2} (B-B_0)-2 \sigma. \label{C3M}
\end{equation}
In our picture, atom scattering in vacuum is mediated by the formation
and decomposition of a collective boson. For the atom phase, one
extracts the scattering length for $k\to 0$ \cite{Diehl:2005ae},
\begin{figure}
\begin{minipage}{\linewidth}
  \begin{center}
\setlength{\unitlength}{1mm}
\begin{picture}(85,50)
      \put (0,0){
     \makebox(80,49){
     \begin{picture}(80,49)
      \put(0,0){\epsfxsize80mm \epsffile{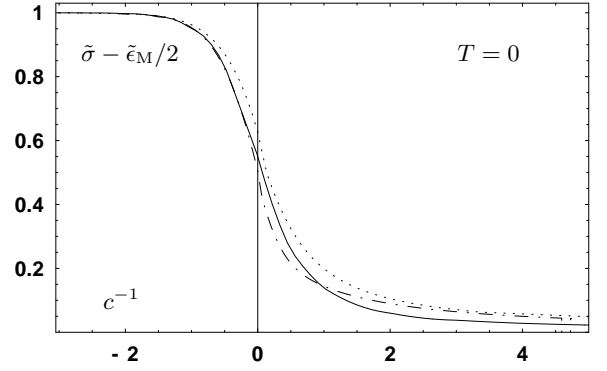}}
      \put(14,8){$c^{-1}$}
      \put(10,41){ $\sigex - \tilde\epsilon_{\text{M}}/2$}
      \put(60,41){ $T=0$}
 \end{picture}
      }}
   \end{picture}
\end{center}
\vspace*{-4.25ex} \caption{Chemical potential 
for $T=0$ minus half the binding energy
  $\tilde\epsilon_{\text{M}}/2 = -\theta(c^{-1}) c^{-2}$.
  We compare our FRG result (solid) to
  extended mean field theory (dotted) and our previous Schwinger-Dyson
  result (dot-dashed) \cite{Diehl:2005ae}. }
\label{MuofInvc}
\end{minipage}
\end{figure}
\begin{equation}
  a=-\frac{\hat{h}_\varphi^2}
{8\pi \hat{k}\hat{m}_\varphi^2 }
  = -\frac{\hat{h}_\varphi^2 \hat{k}}{16\pi M \mu (B-B_0)}. \label{C3A}
\end{equation}
\Eq{C3A} relates $h_{\varphi,\text{in}}^2=\hat{h}_\varphi^2$ to the
scattering length $a(B)$, thus fixing all parameters of our
model. Eq. \eqref{C3A} can also be used for $B<B_0$.  Integrating
Eq.~\eqref{VacFlow} for $\sigma={\sigma}_{\text{A}}<0$ with the condition
$\hat{m}_\varphi^2(k=0)=0$ yields the well-known relation between
molecular binding energy and scattering length
$\epsilon_{\text{M}}={{\sigma}_{\text{A}} \hat{k}^2}/{M} =-{1}/({M a^2}).$

The flow of the renormalised Feshbach coupling $h_\varphi^2$ is
determined by the anomalous dimension,
\begin{equation}
\partial_t \left(\frac{h_\varphi^2}{k}\right) =(-1+\eta_\varphi)
  \frac{h_\varphi^2}{k}, \quad
\eta_\varphi= \frac{h_\varphi^2}{6\pi^2} \frac{k^5}{(k^2-
    {\sigma})^3}.
\end{equation}
For $\sigma=0$, the rescaled renormalised Feshbach coupling
rapidly approaches a fixed-point (scaling solution) given by
$
\eta_\varphi=1,\quad h_\varphi^2/k=6\pi^2.
$
In the vacuum, we find $A_\varphi=1/2$ and
$Z_\varphi(\sigma_{\text{A}}<0,k\to0)= 1+
  {\hat{h}_\varphi^2}/({32\pi
    \sqrt{-\sigma_{\text{A}}}}) $.

Next, we study the equation for the dimensionless four-boson coupling
$\hat \lambda_\varphi = Z_\varphi^2 \lambda_\varphi$,
\begin{equation}
  \partial_t \hat \lambda_\varphi = -
  \frac{\hat h_\varphi^4}{4\pi^2}\frac{k^5}{(k^2 -\sigma)^4} +
  \frac{2\sqrt{2}\hat \lambda_\varphi^2}{3\pi^2}
  \frac{A_\varphi (1-\frac{\eta_\varphi+
      \eta_{A_\varphi}}{5}) k^5}
  {(2 Z_\varphi A_\varphi k^2 + \hat m_\varphi^2)^2}.
       \label{eq:17}
\end{equation}
There are contributions from fermionic \emph{and} bosonic vacuum
fluctuations, but no contribution from higher $\rho$ derivatives of
$u$. For $\sigma=0$ and large $\hat h_\varphi^2$, we use the scaling form
$Z_\varphi=\hat{h}_\varphi^2/(6\pi^2 k)$,
$\hat{m}_\varphi^2=\hat{h}_\varphi^2 k/(6\pi^2)$, $A_\varphi
=\frac{1}{2}$, $\eta_\varphi=1$, $\eta_{A_\varphi}=0$ and find for the
ratio $Q=\hat{\lambda}_\varphi k^3/\hat{h}_\varphi^4$ the flow
equation
\begin{equation}
  \partial_t Q = 3 Q -\frac{1}{4\pi^2} + \frac{3\pi^2
    }{\sqrt{2}}\,  Q^2. \label{AZ31}
\end{equation}
The infrared stable fixed point $Q_* \simeq 0.008$ corresponds to a
renormalised coupling
\begin{equation}
  \lambda_\varphi = \frac{\hat{\lambda}_\varphi}{Z_\varphi^2} =
  \frac{36 \pi^4
    Q_*}{k}, \label{AZ32}
\end{equation}
to be compared with the effective four-fermion coupling
$\lambda_{\psi,\text{eff}}= -\hat{h}_\varphi^2/\hat{m}_\varphi^2 =-
6\pi^2/k$. The constant ratio between these two quantities is the
origin of the universal ratio between the scattering length for
molecules and atoms,
$a_{\text{M}}/a=2\lambda_\varphi/\lambda_{\psi,\text{eff}}$.
\begin{figure}
\begin{minipage}{\linewidth}
  \begin{center}
\setlength{\unitlength}{1mm}
\begin{picture}(85,50)
      \put (0,0){
     \makebox(80,49){
     \begin{picture}(80,49)
       \put(0,0){\epsfxsize80mm \epsffile{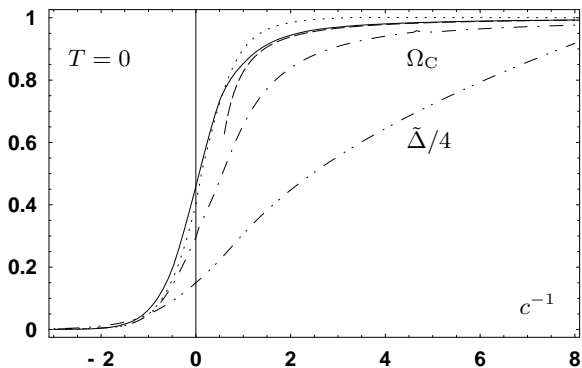}}
       \put(70,8){$c^{-1}$}
       \put(55,30){$\tilde \Delta/4$}
       \put(10,41){$T=0$}
       \put(55,41){$\Omega_{\text{C}}$}
   \end{picture}
      }}
   \end{picture}
\end{center}
\vspace*{-4.25ex} \caption{Condensate fraction $\Omega_{\text{C}}$
  (solid) and gap parameter $\tilde\Delta$ (dash-double-dotted) at
  $T=0$. We compare $\Omega_{\text{C}}$ with extended mean field
  theory (dotted) and Schwinger-Dyson equations (dot-dashed)
  \cite{Diehl:2005ae}. The condensate fraction matches a
  phenomenological Bogoliubov theory with $a_{\text{M}} = 0.92 a$ in
  the BEC regime (dashed), consistent with our vacuum result.}
\label{OmCofT2}
\end{minipage}
\end{figure}

In the molecule phase for $\sigma_{\text{A}}<0$ and $k=0$, one has $
\lambda_{\psi,\text{eff}}= {8\pi}/{\sqrt{-{\sigma}_{\text{A}}}} $
\cite{Diehl:2005ae}.  Omitting the molecule fluctuations, a direct
integration of Eq.~\eqref{eq:17} yields
$\lambda_\varphi=8\pi/\sqrt{-\sigma_{\text{A}}}$ and therefore
$a_{\text{M}}/a=2$, whereas the molecule fluctuations lower this
ratio. With the cut-off functions \eqref{eq:Cutoff} we get
$a_{\text{M}}/a =0.92$, while further optimisation of $R_k$ leads to
$a_{\text{M}}/a =0.71$. Similar diagrammatic approaches give
$a_{\text{M}}/a =0.75(4)$ \cite{AAAAStrinati}, whereas the solution of
the 4-body Schr\"{o}dinger equation yields $a_{\text{M}}/a =0.6$
\cite{Petrov04}, confirmed in QMC simulations \cite{Giorgini04} and
with diagrammatic techniques \cite{Kagan05}.

\emph{Many-body problem} -- The system is now characterised by two
additional scales, the temperature $T$ and the Fermi momentum $\kF$.
We set $\hat{k}=\kF$ from now on and use tildes instead of hats in
order to indicate this specific normalisation. We determine the
initial values for the flow in these units by
\begin{figure}[t!]
\begin{minipage}{\linewidth}
\begin{center}
\setlength{\unitlength}{1mm}
\begin{picture}(85,50)
      \put (0,0){
     \makebox(80,49){
     \begin{picture}(80,49)
       \put(0,0){\epsfxsize80mm \epsffile{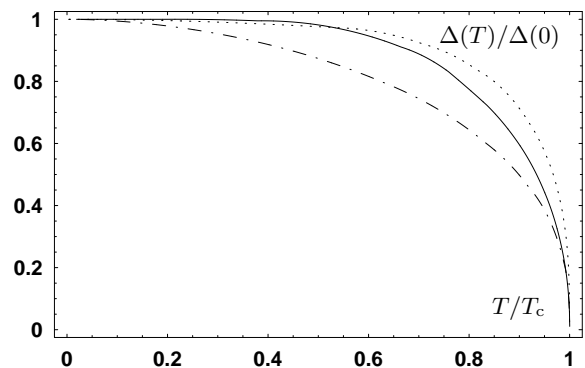}}
       \put(65,8){ $T/T_{\text{c}}$}
       \put(58,44){ $\Delta(T)/\Delta(0)$}
\end{picture}
      }}
   \end{picture}
\end{center}
\vspace*{-4.25ex} \caption{Temperature dependence of the gap
  $\Delta(T)/\Delta(0)$ in the BCS (solid, $c^{-1}=-2$), resonance
  (dotted, $c^{-1}=0$) and BEC regime (dot-dashed, $c^{-1} =4$).}
\label{OmCofT1}
\end{minipage}
\end{figure}
Eqs.~\eqref{C3M},\eqref{C3A} in terms of the concentration $c=a\kF$
and $\tilde{h}_\varphi^2$. For large $\tilde{h}_\varphi^2$ (broad
Feshbach resonance), the value of $\tilde{h}_\varphi^2$ will not be
relevant.  Finally, we have to adjust $\tilde{\sigma}$ in order to
obtain the correct density, which is related to the $\tilde{\sigma}$
dependence of the potential at its minimum. Within our normalisation,
this yields the condition $ {\partial u_{\text{min}}}/{\partial
  \tilde{\sigma}} =- {1}/{(3\pi^2)}$ for $k=0$. We follow the flow of
$\partial u_{\text{min}}/\partial \tilde{\sigma}$ by taking the
$\tilde{\sigma}$ derivative of Eq.~\eqref{PotFlow}, starting with an
initial value $-\tilde\sigma^{3/2} \theta (\tilde \sigma)/(3\pi^2)$ at
$k_{\text{in}}$. The flow equation integrates out the modes around the
Fermi surface for $\tilde\sigma >0$. At least for low $T$, the
different contributions on the right-hand side can be identified with
the densities in unbound atoms, molecules and the condensate
\cite{Diehl:2005ae}. Our result for $\tilde{\sigma}(c^{-1})$ is shown
in Fig. \ref{MuofInvc}. On resonance, we obtain
$\tilde\sigma(c^{-1}=0)=0.55$, while quantum Monte Carlo (QMC)
simulations give $\tilde\sigma(c^{-1}=0)=0.44(1)$ \cite{Carlson03},
$\tilde\sigma(c^{-1}=0)=0.42(2)$ \cite{Giorgini04}.

The density and temperature effects modify the flow when $k\approx 1$
or $k\approx \tilde T^{1/2}$, i.e., when the wavelength of
fluctuations being integrated out is comparable to the interparticle
spacing or the de Broglie wavelength. For $T=0$, in particular,
$m^2_\varphi$ reaches zero for $k_{\rm SSB}>0$, and the flow has to be
continued in the SSB regime with $\rho_0(k<k_{\rm SSB})>0$ until
$k\to0$. We show in Fig.~\ref{OmCofT2} the condensate fraction
$\Omega_{\text{C}}$ \cite{Diehl:2005ae} and the gap for single
fermionic atoms $\tilde\Delta=h_\varphi\sqrt{\rho_0}$. In the BCS
regime, the BCS value
($\tilde\Delta(c^{-1})/\tilde\Delta^{BCS}(c^{-1})=0.9$) for the gap
parameter is approximately reproduced. On resonance, we find
$\tilde\Delta(c^{-1}=0)=0.6$, to be compared to the QMC value
$\tilde\Delta(c^{-1}=0)=0.54$ \cite{Carlson03}.

At higher temperature, the effects of fermionic fluctuations on the
build-up of $\rho_0$ are reduced and the bosonic fluctuations tend to
diminish $\rho_0$. At $T_{\text{c}}$ where $\rho_0(k\to0)\to0$, we
find a second-order phase transition.  The critical region is governed
by boson fluctuations with universal properties in the $O(2)$
universality class. From the scaling solution, we find a critical
exponent $\eta=\eta_\varphi+\eta_{A_\varphi} \approx 0.05$ throughout
the crossover.  We plot $\Delta(T)/\Delta(0)$ for different values of
$c^{-1}$ in Fig.~\ref{OmCofT1}.  The universal behaviour is visible
for $T\to T_{\text{c}}$. On the BCS side, the scale $k_{\rm SSB}$ goes
to zero for $c^{-1}\to -\infty$, leading to an exponentially
suppressed gap.

\begin{figure}[t!]
\begin{minipage}{\linewidth}
\begin{center}
  \setlength{\unitlength}{1mm}
\begin{picture}(85,50)
      \put (0,0){
     \makebox(80,49){
       \begin{picture}(80,49)
         \put(0,0){\epsfxsize80mm
           \epsffile{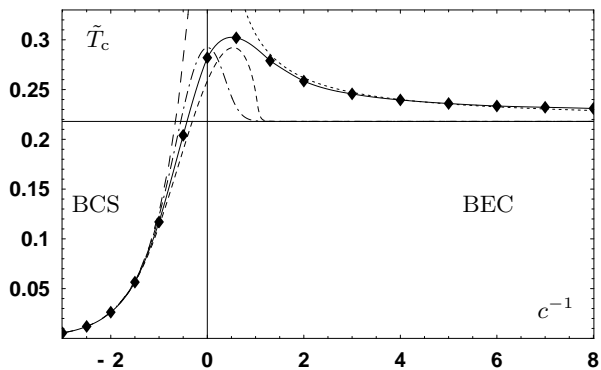}}
         \put(70,8){ $c^{-1}$}
         \put(7,22){ { BCS }}
       \put(60,22){{ BEC}}
       \put(10,44){ $\tilde \Tn_{\text{c}}$ }
\end{picture}
      }}
   \end{picture}
\end{center}
\vspace*{-4.25ex} \caption{Crossover phase diagram from the FRG
  approach (diamonds). We indicate the BCS (long dashed) and
  free BEC (horizontal line) values of $\tilde T_{\text{c}}$ and compare with
  Schwinger-Dyson equations (dashed and dot-dashed) \cite{Diehl:2005ae}.}
\label{PhaseDiag}
\end{minipage}
\end{figure}

The phase diagram in the $(\tilde T,c^{-1})$ plane is shown in
Fig.~\ref{PhaseDiag}. In the regime of weak attractive interactions, the BCS critical temperature is reproduced. On the BEC side, we find the shift of the
critical temperature $\Delta T_{\text{c}}/T_{\text{c}}^{BEC}= \kappa
(n_{\text{M}})^{1/3} a_{\text{M}} =(6 \pi^2)^{-1/3}\kappa
(a_{\text{M}}/a) c$ \cite{Blaizot:2006vr} with $\kappa= 1.7$,
$a_{\text{M}}=0.92 a$ (short dotted line). Lattice simulations give
$\kappa =1.32(2)$ \cite{Arnold:2001mu}. 

In the BEC regime, both at zero temperature and close to $T_{\text{c}}$ the many-body physics reflects the behavior of ``fundamental'' bosons of mass $2M$ interacting via a scattering length $a_{\text M} = 0.9 a$. This demonstrates the emergence of an effective bosonic theory, where all memory of the truly fundamental fermionic constituents has been lost, easily understood by the fact that the binding energy of the molecules is the largest scale in this region of the phase diagram. Moving to the unitary and BCS regimes, only a very narrow region around $T_{\text c}$ is dominated by bosonic fluctuations, which give rise to the above mentioned critical behavior. Bosonic degrees of freedom are, however, crucial to accomodate the symmetry requirements in the spontaneously broken phase throughout the crossover and form an important building block for our evaluation scheme.


Our present truncation does not yet include the effects of particle-hole 
fluctuations. They lead to a strong decrease of both the critical 
temperature and the gap parameter in the regions of the phase diagram 
where there is a substantial Fermi surface, i.e., in the BCS and
unitary regimes. In the BCS regime, the reduction of both $T_c$ and 
$\Delta(T=0)$ by a factor $(4 e)^{1/3} \approx 2.2$ is well known as 
Gorkov's effect \cite{Gorkov61}. In our formulation, this effect is 
encoded in the running of an effective four-fermion vertex which is 
generated by certain mixed boson-fermion diagrams. The relevant diagrams 
precisely have the topology of the particle-hole diagrams in a purely 
fermionic setting. This will be discussed in future work.

\emph{Conclusion} -- Our functional renormalization group analysis for
ultracold fermionic atoms clearly demonstrates the necessity of the
inclusion of bosonic quantum and statistical fluctuations beyond
extended mean field theory.  Both the BEC regime ($c^{-1}\to\infty$)
and the universal critical behaviour ($T\to T_{\text{c}}$) are
dominated by bosons. The vacuum fluctuations are crucial for the
four-boson interaction. The thermal boson fluctuations are needed to
establish the expected second-order phase transition. Our method is
technically simple and involves only a few running couplings, still
enough to resolve the full range of microscopic couplings, i.e., the
BCS-BEC crossover, as well as the whole range of temperatures from the
ground state to the phase transition. We control all regimes of
densities including the physical vacuum ($k_{\text{F}}\to 0$) where
the crossover terminates in a second-order vacuum phase transition.
The simplicity of the picture constitutes an ideal starting point for
systematic quantitative improvements by extending the truncation.  For
example, we have not yet included the (many-body) effect of
particle-hole fluctuations which will lower $T_{\text{c}}$ in the BCS
and crossover regimes.  Extended truncations should lead to
quantitative precision for the crossover physics.

We thank S.~Fl\"orchinger, H.C.~Krahl, M.~Scherer and P.~Strack for useful
discussions.


\end{document}